\def\etal{{\it\ et al.}}

\def\Ha{H$\alpha$}
\def\HI{H${\scriptstyle\rm I}$}
\documentclass[preprint2]{aastex}

\newcommand{\ffffff}[1]{\mbox{$#1$}}

\newcommand{\scmd}{\mbox{\ffffff{''}}}

\def\ltsima{$\; \buildrel < \over \sim \;$}
\def\simlt{\lower.5ex\hbox{\ltsima}}
\def\gtsima{$\; \buildrel > \over \sim \;$}
\def\simgt{\lower.5ex\hbox{\gtsima}}

\slugcomment{To appear in PASP}

\shorttitle{G. F. Lewis\etal}
\shortauthors{Dark Matter in High Velocity Clouds}

\begin{document}

\title{Detecting Dark Matter in High Velocity Clouds}

\author{Geraint F. Lewis\altaffilmark{1,5}, 
Joss Bland-Hawthorn\altaffilmark{1,6},
Brad K. Gibson\altaffilmark{2,3,7} \&
Mary E. Putman\altaffilmark{4,8}}

\altaffiltext{1}{
Anglo-Australian Observatory, P.O. Box 296, Epping, NSW 1710,
Australia}
\altaffiltext{2}{
CASA, University of Colorado, Boulder, CO,
USA}
\altaffiltext{3}{
Astrophysics \& Supercomputing Centre, Swinburne University of Technology,
Mail 31, P.O. Box 218, Hawthorn, Victoria 3122, Australia}
\altaffiltext{4}{
Research School of Astronomy \& Astrophysics, Institute of Advanced Studies,
Australian National University, Mount Stromlo Observatory, Weston, ACT, Australia}

\altaffiltext{5}{
\tt gfl@aaoepp.aao.gov.au
}
\altaffiltext{6}{
\tt jbh@aaoepp.aao.gov.au
}
\altaffiltext{7}{
\tt bgibson@mania.physics.swin.edu.au
}
\altaffiltext{8}{
\tt mary.putman@atnf.csiro.au
}

\begin{abstract}
\noindent Many high velocity \HI\ clouds (HVCs) are now believed to be
scattered  throughout   the  Galactic  halo  on  scales   of  tens  of
kiloparsecs.  Some of these  clouds appear to contain substantial \HI\
masses   ($>10^6$M$_\odot$).   It  has   been  suggested   that  these
structures may  be associated with dark matter  `mini halos' accreting
onto the Galactic  halo.  For a compact HVC along the  sight line to a
more distant galaxy, we demonstrate that `pixel gravitational lensing'
provides a crucial test for the presence of a dark halo in the form of
massive compact  objects. The detection of pixel  lensing will provide
an independent means to map the mass distribution within HVCs.
\end{abstract}


\section{INTRODUCTION}\label{introduction}

Observations  of the  Galactic halo  make a  compelling case  that the
formation of halos continues to  the present day (Wyse 1999). The halo
appears to have built up through a process of accretion and merging of
low-mass  structures  which  is  still   going  on  at  a  low  level.
Hierarchical cold dark matter (CDM) simulations, however, predict that
the Galactic halo  should have many more satellites  than are actually
observed (Klypin\etal\  1999; Moore\etal\ 1999).   Observations reveal
that much of the sky is peppered with high-velocity \HI\ clouds (HVCs)
which  do  not  conform  to  orderly  Galactic  rotation.   These  are
interesting  accretion   candidates  $-$  particularly   if  they  are
associated  with  dark  matter  `mini  halos' $-$  except  that  their
distances, $d$, are  unknown for all but a few  sources.  As a result,
fundamental  physical  quantities  $-$  size ($\propto  d$)  and  mass
($\propto  d^2$)  $-$  are  unconstrained which  has  encouraged  wide
speculation  as to the  nature of  HVCs (Wakker  \& van  Woerden 1997;
1999).

The current  renaissance in HVC studies  can be traced in  part to one
paper.  Blitz\etal\ (1999) have  shown that the velocity centroids and
groupings  of positive/negative  velocity  clouds on  the  sky may  be
understood  within  a reference  frame  centered  on  the Local  Group
barycenter.   They interpret  HVCs as  gas clouds  accreting  onto the
Local  Group over  a megaparsec  sphere. Braun  \& Burton  (2000) have
identified  specific examples  of compact  clouds that  have `rotation
curves' consistent with CDM mass profiles. For sources at 700~kpc, the
kinematic  signatures  imply  a  high dark-to-visible  mass  ratio  of
10$-$50.

However,  Zwaan \&  Briggs (2000)  note  from the  Arecibo \HI\  Strip
Survey that of 300 galaxies and  14 galaxy groups, none appear to have
properties resembling HVCs associated with  the Milky Way or the Local
Group.   A possible  interpretation is  that the  clouds  are somewhat
closer to the galaxy or group barycenters.

Reliable distance indicators  for HVCs are presently hard  to come by.
Oort (1966) proposed virial distances  on the assumption that HVCs are
self-gravitating. The mass inferred from the \HI\ column density has a
different  distance dependence  to the  gravitationally  inferred mass
from  the  \HI\ line  width.   This crude  method  puts  some HVCs  at
megaparsec  distances (Blitz\etal\  1999; Braun  \& Burton  1999).  An
alternative  method  is to  use  bright  sources with  well-calibrated
distances along  the sight line to  gas clouds. If  the cloud produces
absorption in  the spectrum of  one source but  not in the  other, the
cloud  distance can be  bracketed.  This  method establishes  that two
HVCs are within 10~kpc of the Sun (van Woerden\etal\ 1999).

We have recently developed the H$\alpha$ distance method which has the
potential to reach  much greater distances (Bland-Hawthorn\etal\ 1998;
Bland-Hawthorn \& Maloney 1999a,b).  If a gas cloud is optically thick
to ionizing  radiation from a  source with known luminosity,  the \Ha\
flux can  be used to infer  the external field  strength and therefore
the cloud's distance  from the source.  Two groups  have now used this
method to show that many HVCs  are faint or invisible in H$\alpha$ and
appear to be  distributed on scales of tens  of kiloparsecs throughout
the Galactic halo.  Some of these clouds have \HI\ masses in excess of
10$^6$M$_\odot$ (Weiner\etal\  {\it in preparation};  Putman\etal\ {\it 
in preparation}).

In  establishing whether  the HVCs  are truly  candidates  for merging
`mini halos',  their dark matter  content needs to be  determined.  By
considering their  potential gravitational lensing  properties, we now
describe a  test to establish  whether compact HVCs do  indeed contain
such dark matter halos.

\section{GRAVITATIONAL LENSING}\label{model}

\subsection{MASS PROFILE}\label{mass}
Utilizing an extensive sample of numerical simulations, Navarro, Frenk
and White (1997) (NFW) recently  suggested that the density profile of
dark  matter halos  formed within  the framework  of cold  dark matter
cosmology can be simply described as
\begin{equation}
\frac{\rho(r)}{\rho_{crit}} = 
\frac{\delta_s}{(r/r_s)(1+r/r_s)^2}
\label{nfw}
\end{equation}
where $r_s$ is  a scale radius which is related  to the $r_{200}$ (the
radius  of the  halo  within which  the  average is  $200 \times$  the
critical density of the Universe, $\rho_{crit}$), by a `concentration'
$c_s$,   such  that   $r_{200}   =  c_s   r_s$.  The   (dimensionless)
characteristic density, $\delta_s$, is simply related to $c_s$ by
\begin{equation}
\delta_s = \left(\frac{200}{3}\right) 
\frac{c_s^3}{\left[\log{(1+c_s)} - c_s/\left(1+c_s\right)\right]} \ .
\end{equation} 
We adopt  this profile in  describing the dark matter  distribution in
HVCs. For a particular halo  mass, defined as $M_{200} = M(<r_{200})$,
the  concentration $c_s$ depends  on the  halo collapse  redshift, and
hence  the cosmology,  power spectrum  etc.  Assuming  $\Omega_o=1$ \&
$\Lambda_o=0$, we employ  the recipe provided by NFW  to determine the
halo profiles  in a standard  CDM cosmology.  A numerical  routine for
calculating these properties was kindly provided by Prof. Navarro.

Bartelmann (1996)  and Wright \&  Brainerd (2000) have  considered the
gravitational lensing  properties of massive galaxy  clusters with NFW
mass   profiles.   Rather   than  the   radial   density  distribution
(Equation~\ref{nfw}), the  important quantity  in such an  analysis is
the projected surface mass density.  This is given by
\begin{equation}
\kappa(x) = 2 \kappa_s \frac{ f(x) }{ x^2 - 1 }
\label{project}
\end{equation}
where $x =  r / r_s$.  The normalizing factor is  given by $\kappa_s =
\delta_s r_s \rho_{crit} / \Sigma_{crit}$,  where $\Sigma_{crit}$  is 
the critical surface mass density to gravitational lensing;
\begin{equation}
\Sigma_{crit} = \frac{c^2}{4 \pi G} \frac{ D_{os} }{ D_{ol} D_{ls} }
\label{critical}
\end{equation}
where $D_{ij}$ are the angular diameter distances between the observer
$(o)$, lens $(l)$  and source $(s)$. The auxiliary  function $f(x)$ is
given by
\begin{equation}
f(x) = \left\{
\begin{array}{ll}
1 - \frac{2}{\sqrt{1-x^2}} \,arctanh  \sqrt{ \frac{1-x}{1+x} } & (x<1) \\
0                                                              & (x=1) \\
1 - \frac{2}{\sqrt{x^2-1}} \,arctan\; \sqrt{ \frac{x-1}{x+1} } & (x>1) \\
\end{array} \right.
\label{project_anc}
\end{equation}
Defining $g(x) = f(x) /  (x^2-1)$, the normalised surface mass density
is given by;
\begin{equation}
\kappa(x) = 3.33\times10^{-13} \frac{\delta_s}{c_s} r_{200} D_{ol}
\left( 1 - \frac{D_{ol}}{D_{os}}\right) g(x) h^{2}
\label{norm}
\end{equation}
where all distances are in kpc, and it is assumed that only objects in
the local Universe are considered, such that $D_{ls} = D_{os}-D_{ol}$.
The Hubble parameter, $h$, is defined such that $H_o = 100h~km/s/Mpc$.

\begin{table}
\begin{center}
\begin{tabular}{|c|c|c|c|}
\hline
$M_{200}$ & $\delta_s$     & $c_s$ & $r_{200}$ \\ \hline
$10^6$    &$1.66\times10^6$& 40.95 & 1.63      \\
$10^7$    &$1.21\times10^6$& 36.40 & 3.51      \\
$10^8$    &$8.61\times10^5$& 31.95 & 7.57      \\
$10^9$    &$5.87\times10^5$& 27.60 &16.30      \\
\hline
\end{tabular}
\caption{\label{table1} The parameters of the NFW density profile, for 
several  fiducial  values  of   $M_{200}$  (in  units  of  $M_\odot$),
corresponding  to the  potential dark  matter masses  of  HVCs.  While
$\delta_s$ and  $c_s$ are dimensionless,  $r_{200}$ is in  kpc.  These
values were calculated assuming $h=1$.}
\end{center}
\end{table}

\subsection{LENSING PROPERTIES}\label{gravlens}
Armed with these  various tools, we can now  examine the gravitational
lensing properties of dark  matter halos of HVCs. Considering fiducial
masses of $M_{200}=10^6,  10^7, 10^8$ \& $10^9~M_\odot$, corresponding
to the  potential dark  matter masses of  HVCs, we determined  the NFW
parameters; these are summarized in Table~\ref{table1}. Firstly, it is
important to determine whether  the projected surface density of these
dark  matter  halos  is  sufficient to  induce  macrolensing  effects,
resulting in multiple images; in  the absence of strong shearing, this
requires  the  normalized surface  mass  density,  $\kappa$ to  exceed
unity.      An    examination    of     Equations~\ref{project}    and
\ref{project_anc}, reveals that $\kappa(x)$ becomes singular at $x=0$ 
and  must meet  the criterion  for producing  multiple images  at some
radius.  At small $x$, the surface density becomes
\begin{equation}
\kappa(x) \sim 2 \kappa_s \log{\left(\frac{2}{e}\frac{1}{x}\right)} \ .
\label{smallx}
\end{equation}
Considering the parameters in  Table~\ref{table1}, and placing the HVC
at any  reasonable distance, $\kappa(x)$  does not exceed  unity until
very small  (and unphysical) radii.  Therefore, such  halos are unable
to produce observable macrolensing  effects and the dark matter cannot
be  probed  by  looking  for  `image splitting'  of  distant  sources.
However, if the dark matter in  HVC is in the form of compact objects,
these will introduce microlensing variability into the observations of
background  sources.  As  potential source  are  extragalactic, namely
galaxies, the resulting microlensing  is unlike that in the Magellanic
Clouds and  the Galactic  Bulge, where an  individual star is  seen to
brighten  and fade  as  a compact  object  crosses the  line-of-sight.
Rather,  as a  patch  of light  from  a distant  galaxy represents  an
unresolved  population of  stars,  any microlensing  will  be seen  as
against this  smooth background. Hence, microlensing  will be detected
as  fluctuations in surface  brightness over  the source  galaxy.  The
framework of this  `pixel-lensing' was laid down by  Crotts (1992) and
Gould  (1996), and  has recently  proved  successful in  a search  for
compact objects along the lines of sight to M31 and the Galactic Bulge
(Crotts  \&  Tomaney  1996;   Tomaney  \&  Crotts  1996;  Alcock\etal\
1999). Pixel  lensing has also been  proposed as a tool  to search for
both intracluster compact objects  (Gould 1995; Lewis, Ibata \& Wyithe
2000) and  cosmologically  distributed  dark  matter (Lewis  \&  Ibata
2000).

Using   Equation~\ref{norm},  and   considering  the   NFW  parameters
presented in  Table~\ref{table1}, the microlensing  optical depths for
the dark  matter halos can be  calculated.  By assuming  that the halo
lies at a distance  of 100kpc from the Earth, in front  of a source at
3Mpc,  Table~\ref{table2}  presents  the  optical  depths  at  several
angular radii. It is immediately  apparent that the optical depths for
all the  models are  small, in the  regime probed by  the microlensing
searches towards  the Galactic Bulge and halo.   Given the subcritical
value of the  microlensing optical depth seen through  the dark matter
halos associated with the HVCs, the expected number of `pixel-lensing'
events  simply scales as  $\Gamma \propto  \kappa$ (see  Binney 2000).
Hence, in  searching for  HVC compact dark  matter objects  objects, a
simple  test presents  itself;  in monitoring  the surface  brightness
distributions of  galaxies, regions that overlap with  the dark matter
halos of HVCs will display pixel-lensing variability. Moreover, as the
optical depth  increases towards  the center of  the HVC  systems, the
resultant number  of microlensing events should  also increase towards
the  center. Given  the simple  linear scaling  between the  number of
events   and  the   surface  mass   density,  the   identification  of
microlensing of systems  viewed through HVCs will not  only detect the
dark matter component, but will  also provide a (non-kinematic) map of
the dark  matter mass over  tens to hundreds of  arcseconds, depending
upon the mass of the halo. A detailed calculation of the optical depth
distribution for specific HVCs and source galaxies is beyond the scope
of  this current  paper  and  will be  presented  elsewhere. A  simple
estimate  of   the  expected  number   of  events  can  be   found  by
extrapolating the  analysis of Binney (2000), who  determines that for
an optical depth of $\kappa=10^{-6}$ and sources at 50~Mpc, monitoring
with a 4-m class (diffraction limited) telescope will uncover three or
four events  per $10^6$ resolution  elements per week.  For  a similar
source distance,  and a HVC located  at 100~kpc, the  optical depth in
the central regions  of the halos presented in  this paper are greater
than $10^{-6}$, with the $10^9~M_\odot$  exceeding this by a factor of
$\sim50$ over a region 10\scmd\ in radius.  Increasing the distance to
the HVC  will similarly increase  the number of  expected microlensing
events.  This analysis indicates that  if HVCs are enshrouded in halos
of  compact dark  matter, then  a substantial  number  of microlensing
events should be detectable.

One potential  contaminant is microlensing  of the sources  by compact
objects  within  our own  Galactic  halo,  or  that of  the  potential
source. Binney  (2000) recently examined the sky  distribution of halo
microlensing optical  depth to  distant sources within  several models
for the mass distribution of the Galaxy. The optical depth is greatest
in the disk  of the Galaxy, and falls  rapidly with Galactic latitude,
falling below $10^{-6}$ for $|b|\simgt 12^o$. At these higher galactic
latitudes even  the least massive  halo considered in this  paper will
dominate  the microlensing optical  depth along  a line-of-sight  by a
factor of  ten. Similarly, the optical  depth through the  halo of the
source  will  also   be  of  order  $10^{-6}$,  and   show  a  similar
distribution over the galaxy. Microlensing by material in the HVC halo
will enhance this value, and will be spatially correlated with the HVC
core, making in discernible from any intrinsic `self-lensing'.

Recent results from the MACHO  (Alcock et al. 2000) and EROS (Lasserre
et al.  2000) studies towards  the Magellanic Clouds suggest that only
$\sim20\%$ (at most) of the dark  matter halo of the Galaxy resides in
the form of compact objects. If this is the case, and the distribution
of dark matter is universal, then the optical depths presented for the
HVCs would have to be scaled by a similar factor. However, if the dark
matter in the  Galactic halo is clumped on large  scales then our view
to  the   Magellanic  Clouds  may   be  through  a   relatively  empty
region.  Such a  picture, which  is consistent  with  the hierarchical
accretion model discussed in this  paper (see Klypin et al. 1999), may
explain  why  our  view  towards  the Galactic  Bulge  appears  to  be
relatively overdense in MACHOs (Binney, Bissantz
\& Gerhard 2000; Alcock et al. 2000a).

\begin{table}
\begin{center}
\begin{tabular}{|c|c|c|c|}
\hline
$M_{200}$ & 1\scmd           & 10\scmd          & 100\scmd         \\ \hline
$10^6$    &$8.5\times10^{-6}$&$3.8\times10^{-6}$&$5.4\times10^{-7}$\\
$10^7$    &$1.8\times10^{-5}$&$9.9\times10^{-6}$&$2.5\times10^{-6}$\\
$10^8$    &$3.8\times10^{-5}$&$2.3\times10^{-5}$&$8.8\times10^{-6}$\\
$10^9$    &$7.4\times10^{-5}$&$4.9\times10^{-5}$&$2.4\times10^{-5}$\\
\hline
\end{tabular}
\caption{\label{table2} The optical depths of dark matter halos as 
a function of radius for the fiducial masses considered in this paper.
The halo is  placed at a distance  of 100kpc, in front of  a source at
3Mpc. At this distance,  100\scmd\ corresponds to 48.5pc.  The optical
depths can be scaled to other distances using Equation~\ref{norm}.}
\end{center}
\end{table}

\section{PROPOSED EXPERIMENT}\label{sources}
The practical limitations to  the pixel lensing technique described in
Section~2 are, of  course, set primarily by the  availability, or lack
thereof,  of (purported) dark  matter-dominated HVCs  suitably aligned
with background galaxies.  To a 5\,$\sigma$ limiting \ion{H}{1} column
density of 7$\times$10$^{17}$\,cm$^{-2}$,  Murphy et~al.  (1995) claim
an HVC sky  covering fraction of $\sim$37\%.  The  high detection rate
of  high-velocity  Galactic  \ion{Mg}{2}  gas seen  toward  background
quasars  (Savage   et~al.   1993)  implies  a   covering  fraction  of
$\sim$50\%  at   column  densities  of  2$\times$10$^{17}$\,cm$^{-2}$.
Clearly, high-velocity gas does  exist, at some level, along virtually
all  extragalactic sightlines.  Unfortunately,  one needs  to exercise
restraint before  proclaiming that  the pixel lensing  experiment will
therefore be a trivial one.

Of  greatest  concern  is  the  potential  contamination  due  to  the
inclusion  of  ``non-dark matter-dominated  HVCs''  in  the above  sky
covering fractions.   Eliminating large, diffuse, HVCs  from the above
sample  (e.g.   Magellanic  Stream,   Complexes  A,  C,  and  M),  and
restricting  the  analysis  to unresolved  ($<$1\,deg$^2$),  isolated,
HVCs,  immediately  reduces  the  covering  fraction by  a  factor  of
$\sim$50 -- to $\sim$0.7\% --  Blitz \& Robishaw (2000).  An even more
stringent  sampling   was  adopted  by  Braun  \&   Burton  (1999)  in
constructing  their   Compact  High-Velocity  Cloud   (CHVC)  catalog,
resulting  in  a  compilation   of  only  65  candidates.   Subsequent
high-resolution imaging  of a subset  of these CHVCs (Braun  \& Burton
2000) shows  that each is  $\sim$0.2\,deg$^2$ in  areal extent,  for a
total sky covering fraction of  $\sim$0.03\% -- more than three orders
of  magnitude lower  than that  found  by Murphy  et~al. (1995).   The
southern  sample  of  CHVCs  in   the  Braun  \&  Burton  catalog  was
necessarily  limited   to  the  older  \ion{H}{1}   survey  by  Bajaja
et~al. (1985);  this has  since been supplanted  by the  Morras et~al.
(2000) survey (a direct southern analog to the Leiden-Dwingeloo Survey
- Hartmann \&  Burton 1997) and  the \ion{H}{1} Parkes All  Sky Survey
(HIPASS  --  Putman  \&   Gibson  1999a,b).   This  latter  survey  is
particularly attuned to the discovery of  CHVCs, as it is the first of
its kind to sub-Nyquist sample  the southern sky.  A visual inspection
of several random HIPASS data  cubes demonstrates that a factor of two
increase in the number of known southern CHVCs (with \ion{H}{1} column
densities $>$10$^{18}$\,cm$^{-2}$) can be expected.

As Blitz \&  Robishaw (2000) demonstrate, the probability  of a chance
alignment of  a CHVC  with a nearby,  background galaxy  is $\sim1\%$.
Even allowing  for the aforementioned expected increase  in the number
of  catalogued CHVCs,  this probability  will remain  $<$2\%.   On the
other hand,  \it if \rm the  simulations of Klypin  et~al.  (1999) and
Moore  et~al.   (1999) are  correct,  one  might  expect there  to  be
300$-$500 CHVCs in the Local  Group, of which only $\sim$100 have been
accounted for.   Perhaps deeper  \ion{H}{1} surveys will  uncover this
missing  population,   or  perhaps   they  will  only   be  discovered
serendipitously  through studies  of background  quasars.  Regardless,
the  inclusion of this  additional hidden  population of  CHVCs, would
increase the probability of finding a CHVC-background galaxy alignment
to 3$-$5\%.

Obviously,  this  will  be   a  challenging  experiment,  but  not  an
impossible one.   We are initiating  a search through the  HIPASS data
cubes,  in  an attempt  to  uncover  prospective candidates.   Several
possibilities currently  exist, although we  stress this is  neither a
finalized nor  complete list  -- HIPASS~1328$-$30 (Banks  et~al. 1999)
and  ESO~383-G087,   both  in  the  Cen~A  Group,   NGC~3109,  in  the
Antlia-Sextans Grouping,  and NGC55  \& AM0106-382 in  Sculptor, each
have CHVCs  lying within 15$'$-30$'$, yet kinematically  separated, 
from the galaxy.

\section{CONCLUSIONS}\label{conclusions}
Cold  dark matter  models  for the  formation  of universal  structure
predict  that the  Galaxy should  be be  surrounded by  many infalling
`clumps' of material.  While the  Galaxy is accompanied by a number of
dwarf galaxies, they cannot  account for the total expected population
of objects. Recently, it  has been suggested that high-velocity clouds
are  accompanied by  halos of  dark  matter and  hence represent,  and
trace, the missing dark matter population.

In this paper we have demonstrated  that while HVCs are too diffuse to
produce  macrolensing  splitting of  distant  sources,  if their  dark
matter consists of compact object,  then this adds to the microlensing
optical depth. This signal is detectable with a 4m class telescope and
a modest  observing campaign  over a  year, and provide  a map  of the
underlying dark matter distribution.   While chance alignments of HVCs
with more  distant galaxies are not common,  several potential sources
already  present themselves. This  situation is  likely to  improve as
current data  is scanned and  future observations are  undertaken.  We
eagerly await the monitoring of  a `mini halo' candidate HVC along the
sight line to a nearby galaxy.

\end{document}